\documentclass[a4paper,11pt]{article}
\usepackage{pos}
\usepackage{amsmath}
\usepackage{mathtools}
\usepackage{subcaption}
\usepackage{cleveref}
\usepackage{graphicx}
\usepackage{wasysym}

\usepackage{tikz}
\usepackage{tikz-3dplot}
\usetikzlibrary{arrows}
\usetikzlibrary{decorations.markings}
\usetikzlibrary{calc}
\usetikzlibrary{decorations.fractals}
\usetikzlibrary{patterns,angles,quotes}
\usepackage{tikz-3dplot}

\definecolor{mymagenta}{RGB}{200, 0, 100}
\definecolor{myblue}{RGB}{45, 48, 146}

\graphicspath{{plots/}}

\newcommand{\SUTwo}{SU$(2)$}

\newcommand{\matr}[2]{\left(\begin{array}{#1}#2\end{array}\right)}

\title{Digitizing $\mathrm{SU}(2)$ Gauge Fields and What to Look Out for When Doing So}

\author[b]{Tobias Hartung}
\author[a]{Timo Jakobs}
\author[c]{Karl Jansen}
\author[d]{Johann Ostmeyer}
\author[a]{Carsten Urbach}

\affiliation[a]{Helmholtz-Institut für Strahlen- und Kernphysik \&
  Bethe Center for Theoretical Physics,\\ Rheinische Friedrich-Wilhelms-Universität Bonn, Nussallee 14-16, 53115 Bonn, Germany}
\affiliation[b]{Northeastern University - London, Devon House, St Katharine Docks, London, E1W 1LP, United Kingdom}
\affiliation[c]{NIC, DESY Zeuthen, Platanenallee 6, 15738 Zeuthen, Germany}
\affiliation[d]{Department of Mathematical Sciences,
	University of Liverpool, United Kingdom}

\emailAdd{timojakobs@uni-bonn.de}

\abstract{
  With the long term perspective of using quantum computers and tensor networks for lattice gauge theory simulations, an efficient method of digitizing gauge group elements is needed. We thus present our results for a handful of discretization approaches for the non-trivial example of SU(2), such as its finite subgroups, as well as different classes of finite subsets. We focus our attention on a freezing transition observed towards weak couplings. A generalized version of the Fibonacci spiral appears to be particularly efficient and close to optimal.
}

\newcommand{\ri}{\mathrm{i}}

\FullConference{%
The 39th International Symposium on Lattice Field Theory,\\
8th-13th August, 2022,\\
Rheinische Friedrich-Wilhelms-Universität Bonn, Bonn, Germany
}

\begin{document}

\maketitle

\hypertarget{introduction}{%
	\section{Introduction}\label{introduction}}

\noindent Hamiltonian formulations of lattice gauge theories promise numerous advantages
over the commonly used Lagrangian Monte Carlo methods. With rapid developments in tensor network simulations, and the ever-growing number of qbits in current quantum devices, it seems ever more likely that such formulations can be efficiently simulated in the not too distant future. In order to implement such simulations, one however needs some form of digitization of the gauge group. In this proceeding, we will investigate such digitizations for the simple non-trivial example of \SUTwo. The methods and results presented can also be found in greater detail in \cite{Hartung2022}.

\hypertarget{su2-partitionings}{%
	\section{SU(2) Partitionings}\label{su2-partitionings}}
\noindent The first thing that comes to mind when discretizing \SUTwo\ are its finite
subgroups. In particular, we will consider the binary tetrahedral group
$\overline{T}$, the binary octahedral group $\overline{O}$ and the binary
icosahedral group $\overline{I}$, with $24$, $48$ and $120$ elements,
respectively~\cite{Klein:1880}. Their elements are evenly distributed across the whole group,
and research on their behaviour as a substitute for \SUTwo\ has already been conducted~\cite{Petcher:1980cq}.

While these subgroups match the full gauge group's behaviour at smaller
value of the inverse coupling $\beta$,
a so-called freezing transition is observed towards $\beta \rightarrow \infty$. The
root cause of this transition is that for a given element no arbitrarily
close neighbours are contained in the subgroup. This puts a lower bound on the
change in action $\Delta S$, caused by the available updates to the gauge
configuration. Typical Monte Carlo algorithms will thus reject almost all
proposed updates to the gauge configuration for a high enough value of
$\beta$. The gauge configuration is thus \emph{frozen} at the state minimizing
the action.

The critical value of $\beta$ beyond which this behaviour is observed will be
referred to as $\beta_c$. Its value increases with the distance of neighbouring
elements, and thus with the number of elements in the subgroup. As however no
arbitrarily large subgroup of \SUTwo\ is available, one needs to resort to sets
of elements which do not form a  subgroup of \SUTwo, but which lie
asymptotically dense and are as isotropically as possible distributed in the
group. We will call these sets \emph{partitionings} of \SUTwo.

Our main strategy for constructing partitionings will be to use the isomorphy
between SU$(2)$ and the sphere $S_3$ in four dimensions, which is defined by
\begin{equation}
	x \in S_3\ \Leftrightarrow\
	\begin{pmatrix}
		x_0 + \ri x_1  & x_2 + \ri x_3 \\
		-x_2 + \ri x_3 & x_0 - \ri x_1 \\
	\end{pmatrix}\in \mathrm{SU}(2)\,. \label{eq:su2S3}
\end{equation}
This approach can also be generalized for other gauge groups of interest.
In fact, any unitary or special unitary
group U$(N)$ and SU$(N)$ can similarly be reduced to a product of spheres.
Further arguments on why this is the case can be found in the article
associated with this proceeding~\cite{Hartung2022}. For now, this just means,
that we are most interested in partitioning schemes that generalise to spheres
of arbitrary dimension.

\subsection{Linear Partitionings}

\begin{figure}
	\centering
	$\vcenter{\hbox{\includegraphics[width=0.22\textwidth]{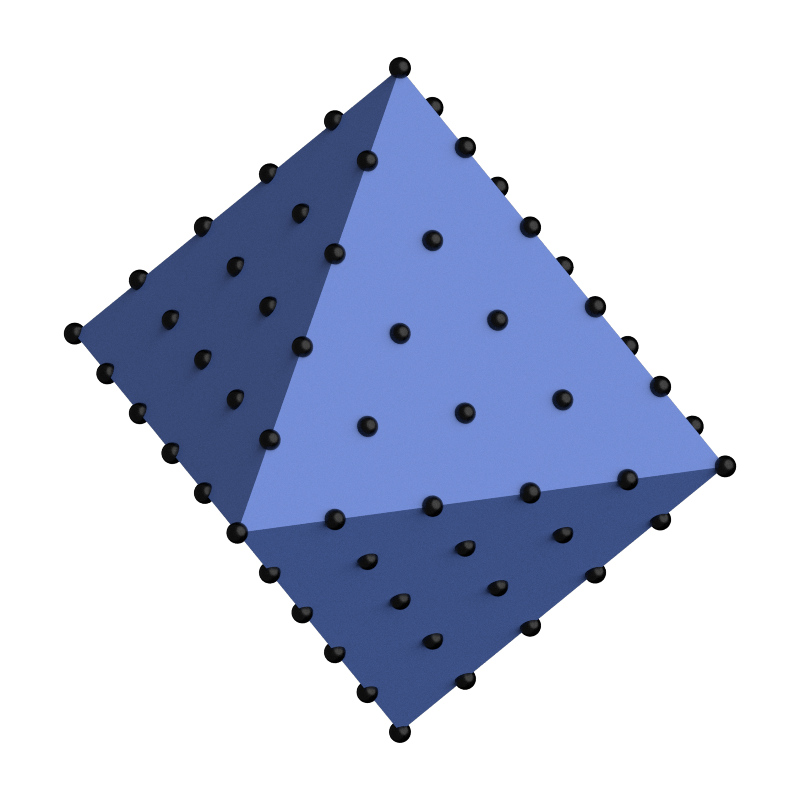}}}
		\vcenter{\hbox{{\huge $\quad \longrightarrow \quad$}}}
		\vcenter{\hbox{\includegraphics[width=0.22\textwidth]{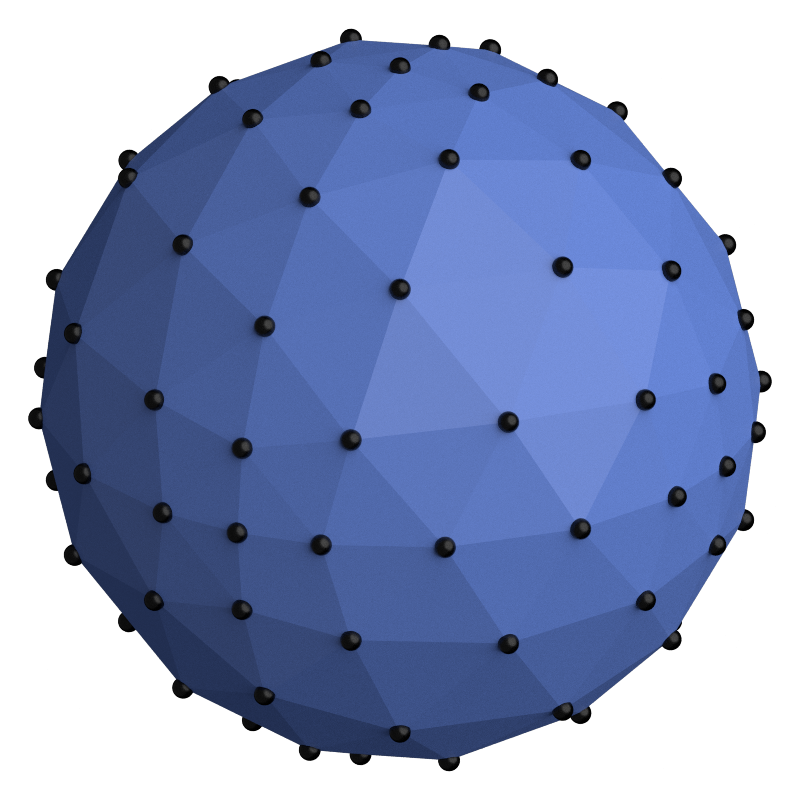}}}$
	\caption{Construction of the linear lattice on $S_2$ for $m=3$. First we
		fill the faces of the octahedron, and then project them out onto the
		sphere. It can be seen that the points are more densely distributed
		around the corners of the original octahedron, and more spread out in the middle of the faces.}
	\label{fig:linLatImg}
\end{figure}

\noindent We begin with the linear partitioning $L_m(k)$ on the sphere $S_k$ in $k+1$
dimensions, given by
\begin{align}
	L_m(k) & \coloneqq \left\{\frac{1}{M}\left(s_0j_0,\dots,s_kj_k\right) \middle|\,\sum_{i=0}^kj_i=m,\;\forall i\in \{0,\dots,k\}:\,s_i\in\{\pm1\},\,j_i\in\mathbb{N}\right\}\,,
	\label{eq:linear_discretisation}
\end{align}
with
\begin{align}
	\qquad M \coloneqq \sqrt{\sum_{i=0}^kj_i^2}\,.\label{eq:define_M}
\end{align}
As depicted in \cref{fig:linLatImg}, it can be visualized as distributing
points evenly on higher dimensional versions of the octahedron, and then
projecting them onto the sphere. Note that some testing on this partitioning
has already been carried out in \cite{Hackett:2018cel}.

One apparent problem with this approach is that group elements are
distributed more densely around the vertices of the octahedron, and more
sparsely in the middle of the faces. As one might expect, this will cause
systematic deviations from the full gauge group, if not corrected for.

To counteract such effects, we assign a probabilistic weight attributed to each
point. This weight $w$ is defined as the volume of the Voronoi
cell~\cite{Voronoi:1908a,Voronoi:1908b} of the point using the canonical
metric on $S_k$ derived from the Euclidean distance. For the linear partitionings this can be approximated with
\begin{align}
	\label{eq:linweights}
	w \approx \left(\frac{\sqrt{2}}{M}\right)^k.
\end{align}
Further details of this calculation can again be found in~\cite{Hartung2022}.

\subsubsection{Volleyball Partitionings}

\noindent Similarly to the linear lattices, we can also construct a partitioning by
distributing points on the faces of a hypercube. On $S_2$ the result looks
somewhat reminiscent of a Volleyball, which is why we will refer to this
approach as the volleyball partitioning:
\begin{align}
	\begin{split}
		V_m(k) &\coloneqq \left\{ \frac{1}{M} \left(s_0 j_0, \dots, s_k j_k  \right) \middle|\,  (j_0, \dots, j_k ) \in \left\{ \text{all perm. of } (m, a_1, \dots, a_k) \right\} , \right.  \\
		& \hspace{8cm} \left. \,s_i\in\{\pm1\}, \, a_i \in \{0, \dots, m\} \vphantom{\frac 1M}\right\}
	\end{split}
\end{align}
where $M$ again defined in \cref{eq:define_M}. Furthermore, we can also include
the trivial case of not subdividing the faces of the hypercube at all:
\begin{align}
	V_0(k) & \coloneqq \left\{\frac{1}{\sqrt{k+1}} \left(s_0, \dots, s_k \right) | \,s_i\in\{\pm1\}\right\}\,.
\end{align}
Similarly to the linear partitionings, points are again distributed more densely towards the corners of the hypercube. In fact, the weights turn out
to be quite similar and are given by
\begin{align}
	\label{eq:Vweights}
	w \approx \left(\frac{1}{M}\right)^k.
\end{align}
For details, we again refer to \cite{Hartung2022}.

\subsubsection{Fibonacci Partitionings}

\begin{figure}
	\centering
	$\vcenter{\hbox{\includegraphics[width=0.22\textwidth]{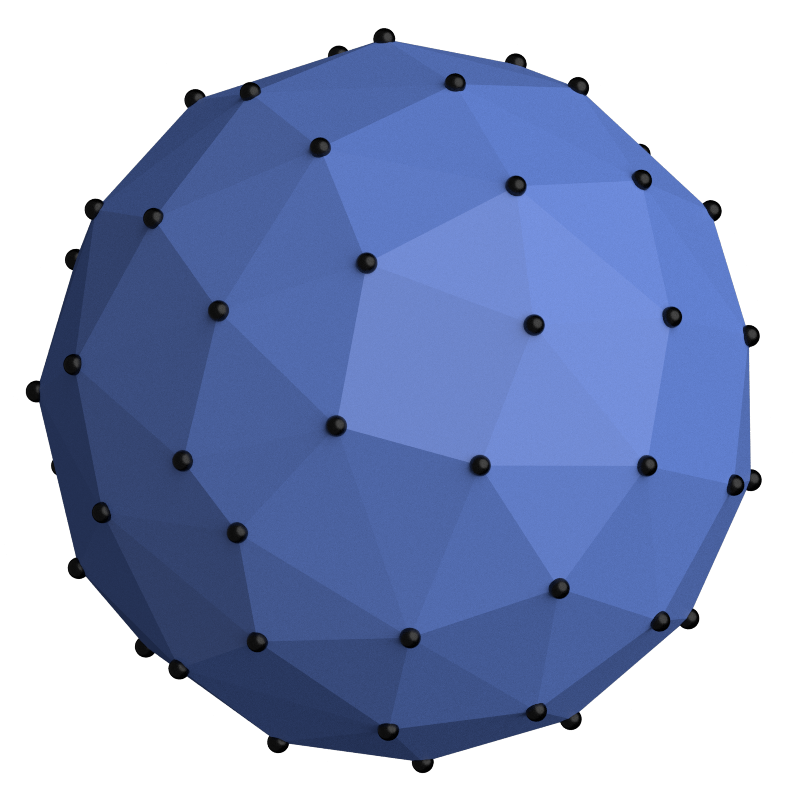}}} \vcenter{\hbox{{\huge $\quad \quad \quad$}}}
		\vcenter{\hbox{\includegraphics[width=0.22\textwidth]{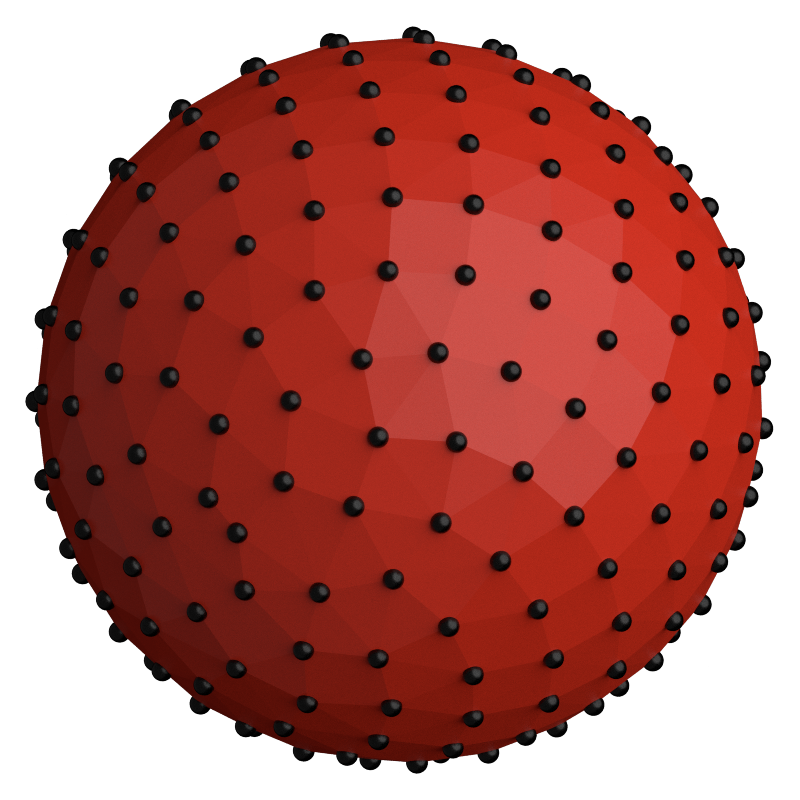}}}$
	\caption{Fibonacci lattices with $64$ lattice points on the left, and $256$ lattice points on the right. It can be seen that the points are spread evenly over the whole sphere, but lack any local geometric structure.}
	\label{fig:fibLatImg}
\end{figure}

The final partitioning of \SUTwo\ considered in this work is a
generalized version of the so-called Fibonacci lattice. For a given size $n$, it is usually constructed within a unit square $[0,1)^2$ as
\begin{align}
	\Lambda_n^2                      = \left\{ \tilde{t}_m \middle| 0 \le m < n, \, \, m \in \mathbb{N} \right\} \qquad \text{with} \qquad \tilde{t}_m & = \begin{pmatrix}x_m\\y_m\end{pmatrix} = \left(\frac{m}{\tau} \quad \mathrm{mod} \quad 1, \frac{m}{n} \right)^t\,, \label{eq:fib2d} \\
	\text{and} \,\,\, \qquad \tau                                                                                                                      & = \frac{1+\sqrt{5}}{2}\,.
\end{align}
Typically it is then mapped onto other surfaces, such as spheres or circles. Some examples
for $S_2$ can be seen in \cref{fig:fibLatImg}. One of the main advantages of
this approach, is that we have precise control over the number of elements in
our partition. This contrary to the case of linear and volleyball partitionings where
the lattice size scales roughly as $n \sim m^3$, and leaves gaps in between the available sizes.
Furthermore, the distribution of points is uniform for larger lattices. Thus
no weighting of the lattice points is needed, and we can set $w=1$. When taking
a closer look at \cref{fig:fibLatImg}, we can however see that the local
structure of this partitioning is fairly irregular. As seen later, this will cause systematic deviations for smaller values of $\beta$.

To construct a Fibonacci partitioning on a $k$-dimensional manifold, we begin by generalizing \cref{eq:fib2d} to the hypercube $[0,1)^k$ embedded in $\mathbb{R}^k$:
\begin{align*}
	\Lambda_n^k & = \left\{ t_m \middle| 0 \le m < n, \, \, m \in \mathbb{N} \right\} \\
	t_m         & = \begin{pmatrix} t_m^1 \\ t_m^2 \\ \vdots \\ t_m^k \end{pmatrix} = \begin{pmatrix}
		\frac{m}{n}        &                      \\
		a_1 \, m \quad     & \mathrm{mod} \quad 1 \\
		\vdots             &                      \\
		a_{k-1} \, m \quad & \mathrm{mod} \quad 1 \\
	\end{pmatrix}
\end{align*}
with
\[
	\frac{a_i}{a_j} \notin \mathbb{Q} \quad \textrm{for} \quad i \neq j \textrm{,}
\]
where $\mathbb{Q}$ denotes the field of rational numbers. The square roots of the prime numbers provide a simple choice for the constants $a_i$:
\begin{align*}
	(a_1, a_2 ,a_3, \dots) = (\sqrt{2}, \sqrt{3}, \sqrt{5}, \dots)
\end{align*}
Then a volume preserving map to the manifold needs to be constructed.  This can be done by studying the jacobian obtained by introducing spherical coordinates on $S_3$~\cite{Hartung2022}. The result is given by
\begin{align*}
	F_n & = \left\{ z\left(\psi_m(t_m^1), \theta_m(t_m^2), \phi_m(t_m^3)\right)  \middle| \, 0 \le m < n, \, \, m \in \mathbb{N} \right\}\,,
\end{align*}
with spherical coordinates on $S_3$
\begin{equation}
	z (\psi, \theta, \phi) = \matr{l}{
		\cos \psi\\
		\sin \psi \cos \theta\\
		\sin \psi \sin \theta \cos \phi\\
		\sin \psi \sin \theta \sin \phi}
\end{equation}
and
\[
	\begin{split}
		\psi_m(t_m^1) & =  \Phi_1 \left( t_m^1 \right) = \Phi_1 \left(
		\frac{m}{n}\right)\,,\\
		\theta_m(t_m^2) & =  \Phi_2\left(t_m^2 \right) =
		\cos^{-1}\left(1-2(m\sqrt{2}\mod 1)\right)\,, \\
		\phi_m (t_m^3) & =  \Phi_3( t^3_m) = 2 \pi (m\sqrt{3} \mod 1)\,.
	\end{split}
\]

\hypertarget{methods}{%
	\section{Methods}\label{methods}}
\noindent With the construction done, it is now time to investigate the behavior
of the proposed partitionings.
For this, we work on a hyper-cubic, Euclidean lattice with the set of lattice sites
\[
	\Lambda\ =\ \{n=(n_0,\ldots, n_{d-1})\in\mathbb{N}_0^d: n_\mu = 0, 1,
	\ldots, L-1\}\,,
\]
with $L\in\mathbb{N}$. At every site there are $d\geq 2$ link variables
$U_\mu(n)\in\mathrm{SU}(2)$ connecting to sites in forward direction
$\mu=0, \ldots , d-1$. We define the plaquette operator as
\begin{equation}
	P_{\mu\nu}(n)\ =\ U^{~}_\mu(n) U^{~}_\nu(n+\hat\mu)
	U^\dagger_\mu(n+\hat\nu) U^\dagger_\nu(n)\,,
\end{equation}
where $\hat\mu\in\mathbb{N}_0^d$ is the unit vector in direction
$\mu$. In terms of $P_{\mu\nu}$ we can define Wilson's lattice
action~\cite{Wilson:1974sk}
\begin{equation}
	S = -\frac{\beta}{2}\sum_n \sum_{\mu<\nu} \mathrm{Re}\,\mathrm{Tr}\,P_{\mu \nu}(n)\,,
\end{equation}
with $\beta$ the inverse squared gauge coupling. We will use the
Metropolis Markov Chain Monte Carlo algorithm to generate chains of
sets $\mathcal{U}_i$ of link variables $\mathcal{U} = \{U_\mu(n):
	n\in\Lambda, \mu=0,\ldots, d-1\}$ distributed according to
\begin{equation}
	\mathbb{P}(\mathcal{U})\ \propto\ \exp[-S(\mathcal{U})]\,.
\end{equation}
This is implemented by iterating over all lattice sites $n$ and all
directions $\mu$. For each pair, we then
\begin{enumerate}
	\item generate a proposal $U_\mu'(n)$ from $U_\mu(n)$.
	\item compute $\Delta S = S(U_\mu'(n)) - S(U_\mu(n))$.
	\item accept with probability
	      \begin{equation}
		      \label{eq:Pacc}
		      \mathbb{P}_\mathrm{acc} = \min\left\{1,\ \exp(-\Delta
		      S)\frac{w(U_\mu'(n))}{w(U_\mu(n))}\right\} \,.
	      \end{equation}
\end{enumerate}
This procedure is repeated $N_\mathrm{hit}=10$ times per
$n$ and $\mu$ before moving on to the next pair $(n, \mu)$.\\

\noindent For the partitionings of \SUTwo\ the proposal $U_\mu'(n)$ is
randomly picked from the neighbors of $U_\mu(n)$. In the case of the linear and volleyball partitionings we will test both the calculated, and trivial ($w=1$) probabilistic weights. As the distribution of points
in the Fibonacci lattices is approximately uniform, only trivial weights are used here.

The full gauge group is approximated by four double precision floating
point numbers, storing $x_i$ from \cref{eq:su2S3}. To stay in \SUTwo\ we
furthermore, restrict $\sum_i x_i^2=1$. New proposals
$U_\mu'(n)$ are generated by multiplying with a random element
$V \in \mathrm{SU}(2)$ which has a maximum distance $\delta$ to the
identity element $1 \in \mathrm{SU}(2)$. $\delta$ controls the average
magnitude of $\Delta S$ in step 2 of the algorithm and is tuned towards
an acceptance rate of about $50\%$ in step 3. As we can pick $\delta$
arbitrarily small (within
the limits of floating point precision), we can avoid the aforementioned
freezing transition. The results obtained with this quasi continuous representation of \SUTwo\ will be referred to as reference
results in the following. \\

\noindent The main observable we will study in this proceeding is the plaquette
expectation value defined as
\begin{equation}
	\langle P\rangle = \frac{1}{N}\sum_{i=1}^N\ P(\mathcal{U}_i)
\end{equation}
with
\[
	P(\mathcal{U}) = \frac{2}{d(d-1) L^d}\sum_n \sum_{\mu<\nu}
	\mathrm{Re}\,\mathrm{Tr}\,P_{\mu \nu}(n)\,.
\]
All results presented here, were obtained in  $3+1$ dimensions, with
$L^4=8^4$ lattice volume. The starting configuration was chosen to be either uniform (cold), or fully random (hot).
Statistical errors for $P$ are computed based on the so-called $\Gamma$-method detailed in Ref.~\cite{Wolff:2003sm} and implemented in the publicly available software package hadron~\cite{hadron:2020}.

\hypertarget{results}{%
	\section{Results}\label{results}}

\begin{figure}
	\centering
	\begin{subfigure}[c]{.49\textwidth}
		\includegraphics[scale=0.75]{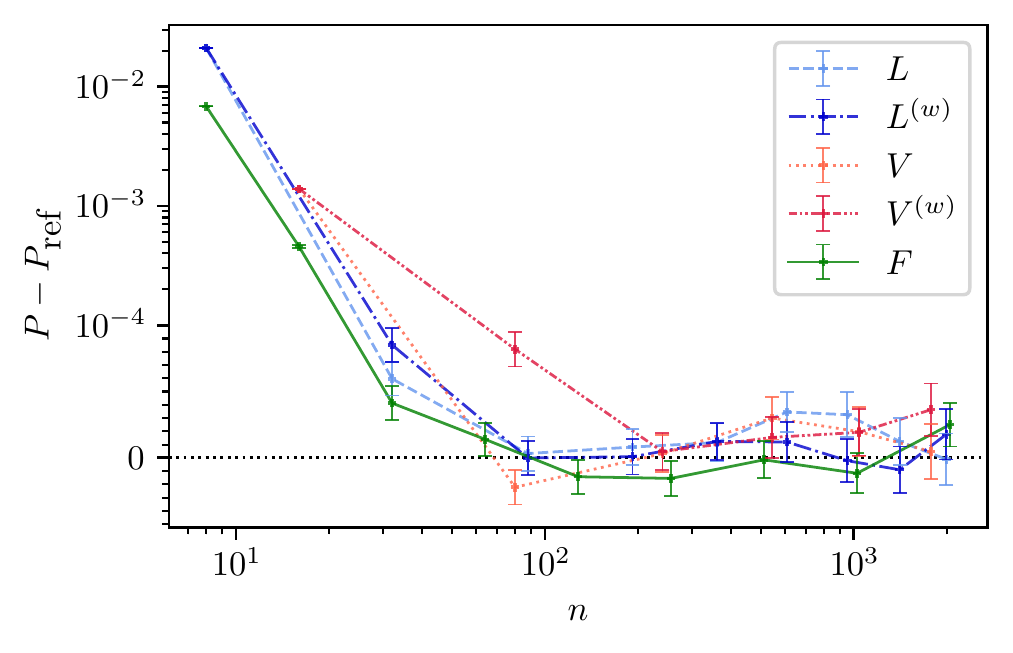}
		\caption{$\beta=1.0$}
	\end{subfigure}
	\hfill
	\begin{subfigure}[c]{.49\textwidth}
		\includegraphics[scale=0.75]{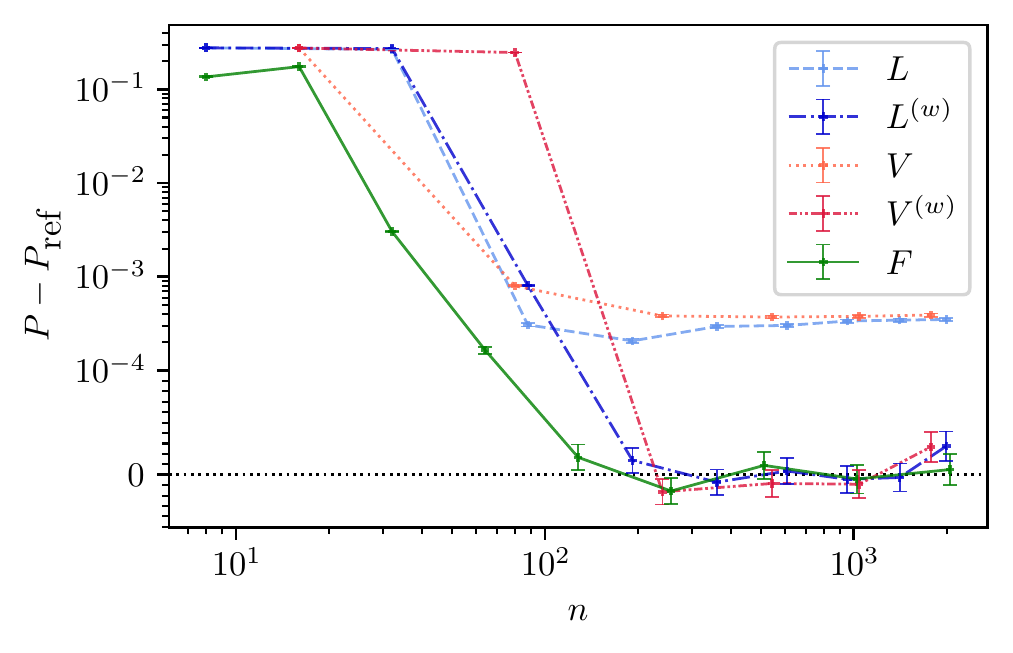}
		\caption{$\beta=3.0$}
	\end{subfigure}
	\caption{Difference between partitioning and reference plaquette expectation value, as a function of the partitioning size $n$. Linear and volleyball partitionings are shown with and without weighting. The partitioning data was averaged over $10^5$, and the reference data over $10^6$ sweeps of the lattice.}
	\label{fig:systematicDeviations}
\end{figure}

\noindent First we want to investigate, how well our partitionings reproduce the reference data, when not affected by the freezing 
transition. For this $3000$ thermalization sweeps, followed by $10^5$ measurement sweeps were performed once for 
$\beta=1.0$ and then again for $\beta=3.0$. This data was then compared to reference data obtained over $10^6$ 
measurement sweeps. The results can be found in \cref{fig:systematicDeviations}.
For $\beta=1.0$ both unweighted and weighted linear and volleyball partitionings mostly reproduce the reference case. 
Noteworthy exceptions are $L_1$ and $V_0$, which are already past their freezing transition, as well as $V_1$ and $L_2$ 
which are pretty close to theirs. For the Fibonacci partitionings no significant deviation is observed for sizes $n \geq 
64$. For $\beta=1$ $F_8$ is again past its freezing transition. For $F_{16}$ and $F_{32}$ the transition is only 
expected for significantly bigger values of $\beta$. A likely explanation for the deviation is thus the irregular 
structure of the partitioning.

At $\beta=3.0$ the picture changes. We now see a clear necessity for considering the Voronoi weights for the linear and volleyball lattices. Without these weights, we see a persistent significant deviation from the reference data.
It is worth noting, that this deviation seems to be mostly independent of the partitioning size. This confirms that it is indeed caused by the uneven density of elements in the partition.

The results for the Fibonacci lattices look similar to the ones at $\beta=1.0$, but shifted to the right. Agreement with the reference data is now achieved for $n \geq 128$. The deviations of the smaller partitionings are most likely to blame on their now much closer freezing transitions.

\subsection{Freezing Transition}

\begin{figure}
	\centering
	\begin{subfigure}[c]{.49\textwidth}
		\includegraphics[scale=0.75]{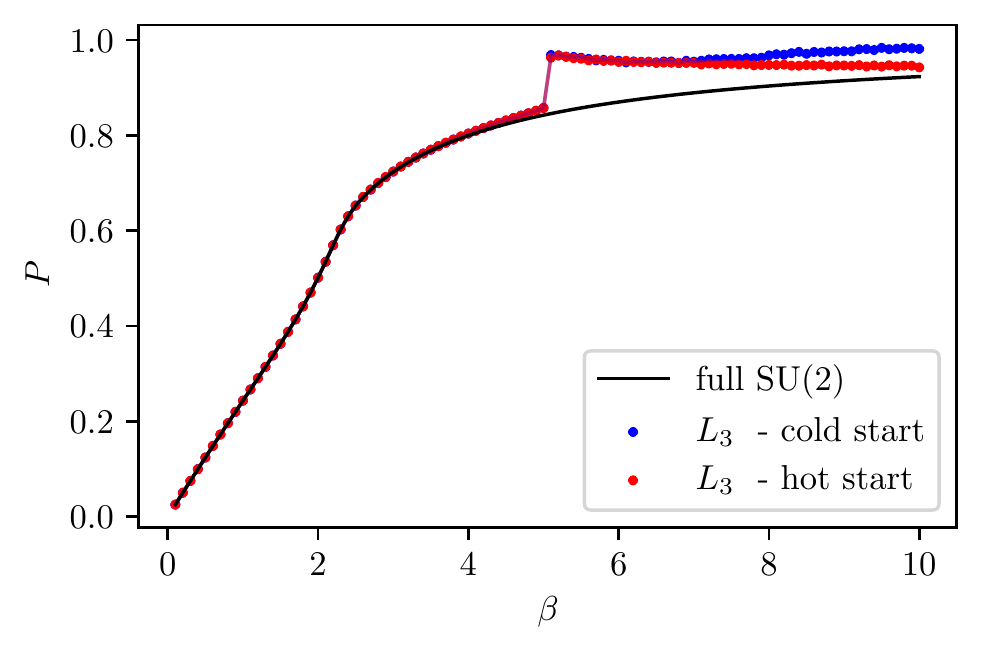}
	\end{subfigure}
	\hfill
	\begin{subfigure}[c]{.49\textwidth}
		\includegraphics[scale=0.75]{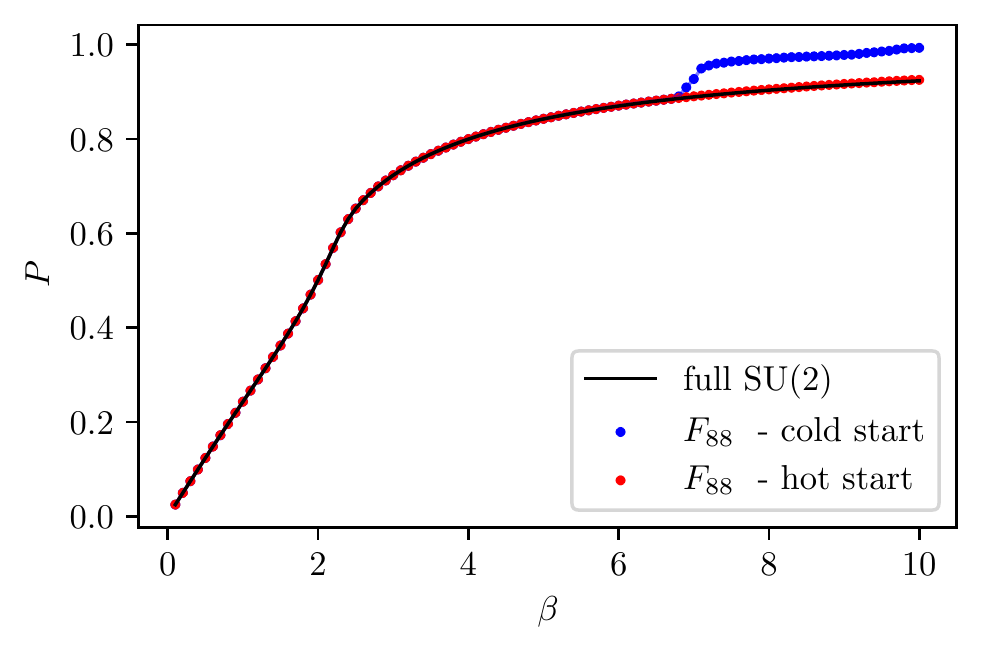}
	\end{subfigure}
	\caption{Hysteresis loops for the Fibonacci partitioning $F_{88}$
		and the linear partitioning with weights included $L_3$. Both have
		$n=88$ elements.}
	\label{fig:betaScanOverview}
\end{figure}
\noindent To study this freezing transition in more detail, we look at $\beta \in \{ 0.1,0.2,\dots,9.9,10.0\}$. For each value of $\beta$, $7000$ sweeps are performed, once with a hot, and once with a cold starting configuration. The plaquette is then measured by averaging over the last $3000$ iterations.

Such scans in $\beta$ can be found in \cref{fig:betaScanOverview}.
$\beta_c$ is then estimated to be the last value before a significant
jump in $\langle P \rangle$, or a significant disagreement between
the hot and cold start\footnote{Fibonacci lattices
    usually show the latter behavior, which is why we increase the
    number of thermalization sweeps to $10^5$. This marginally
    raised the values of $\beta_c$.}. We have checked that the
such determined critical $\beta$-values do not depend significantly
on the volume.

\begin{figure}
    \centering
    \includegraphics[scale=0.75]{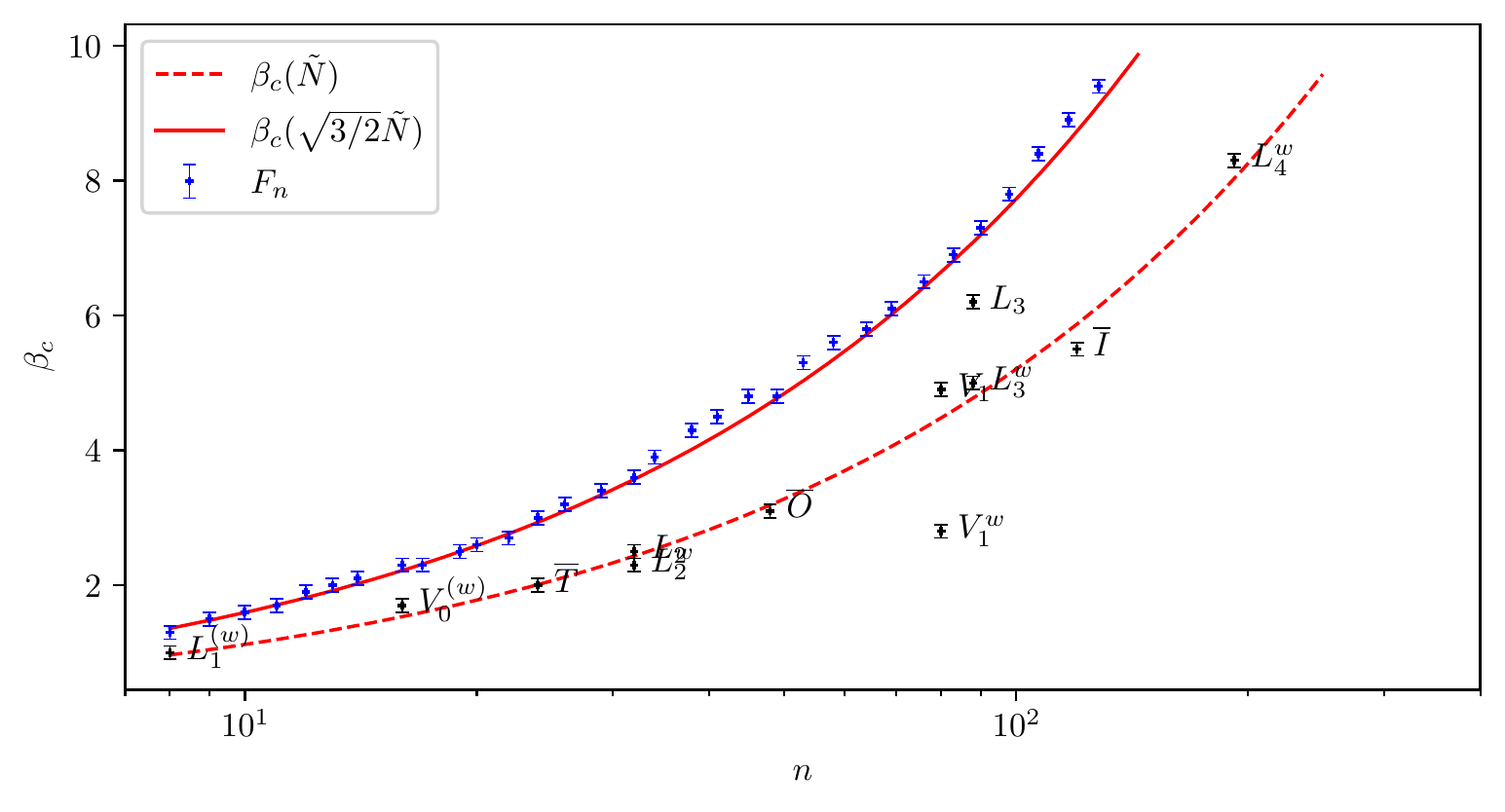}
    \caption{The critical value $\beta_c$ as a function of the number
        $n$ of elements in the set. The lines represent the
        approximation~\cref{eq:betac} where the order $\tilde{N}(n)$ is obtained
        from \cref{eq:approx_order}.}
    \label{fig:fibPhaseScan}
\end{figure}

In \cref{fig:fibPhaseScan} we show the $\beta_c$-values at which the
freezing transitions takes place as a function of the number $n$ of
elements in the set of points or the subgroups. It can be seen that the Fibonacci lattices outperform all other partitionings, and thus offer the biggest range of beta for a given size $n$. Furthermore, it seems that $\beta_c$ is smaller when considering the Voronoi weights for the linear and volleyball lattices, compared to the trivial weights.\\

\noindent In Ref.~\cite{Petcher:1980cq} the authors find that the critical
$\beta$-value can be computed theoretically, at least approximately,
for the finite subgroups:
\begin{equation}
    \label{eq:betac}
    \beta_c(N) \approx \frac{\ln\left(1+\sqrt{2}\right)}{1-\cos(2\pi/N)}\,.
\end{equation}
Here $N$ denotes the power to which the closest neighbors $g$ of the
identity need to be taken, such that $g^N = 1$. Fibonacci, linear and
volleyball partitionings however lack group structure. Therefore, it is not
guaranteed that there is an $N\in\mathbb{N}$ for which $g^N=1$.

Thus, we have to approximate the order $N$. For (approximately)
isotropic discretisations such as the finite subgroups and the
Fibonacci partitioning a global average over the point density is
bound to yield a good approximation for the elements in $C(G)$ and
therefore $N$. The volume of the three-dimensional unit sphere is
$2\pi^2$. If we then assume a locally primitive cubic lattice,
the average distance of $n$ points in $S_3$ becomes
\begin{equation}
    d(n) = \left(\frac{2\pi^2}{n}\right)^{1/3}\,.\label{eq:average_distance}
\end{equation}
Two points of this distance together with the origin form a triangle
with the opening angle
\begin{equation}
    \alpha(n) = 2\arcsin \left( \frac{d(n)}{2} \right)\,,
\end{equation}
thus, a first approximation of the cyclic order is obtained by
\begin{equation}
    \label{eq:approx_order}
    \tilde{N}(n) = \frac{2\pi}{\alpha(n)}\,,
\end{equation}
which solely depends on the number $n$ of elements in the partition.

Note that the assumption of a primitive cubic lattice is even asymptotically incorrect for all the partitionings discussed in this work, and at best a good approximation. How good an approximation it is, can only be checked numerically. In specific cases, it needs further refinement.

In particular, in the case of the Fibonacci partitioning the approximation has
to be adjusted. Since the points are distributed irregularly in this
case, a path going around the sphere does not lie in a two-dimensional
plane. Instead, it follows some zigzag route which is longer than the
straight path. Assuming the optimal maximally dense packing, we expect
the points to lie at the vertices of tetrahedra locally tiling the
sphere. The length of the straight path would then correspond to the
height of the tetrahedron, whereas the length of the actual path
corresponds to the edge length. Their ratio is $\sqrt{3/2}$, so
$\tilde N$ has to be rescaled by this factor to best describe
$\beta_c$ for the Fibonacci partitioning.

We show the curve~\cref{eq:betac} using $\tilde{N}(n)$ and
$\sqrt{3/2}\tilde{N}$, respectively, in addition to the data in
\cref{fig:fibPhaseScan}. The version with $\tilde{N}$ is in very good
agreement with the results obtained for the finite subgroups, while the
rescaled version matches the values for the Fibonacci partitioning
remarkably well.

\hypertarget{conclusion}{%
    \section{Conclusion}\label{conclusion}}

\noindent In this proceeding, we have presented several asymptotically dense
partitionings of SU$(2)$, which do not represent subgroups of SU$(2)$
but which have adjustable numbers of elements. MC simulations restricted to these partitionings turned out to give good agreement with the standard simulation code, in particular when correcting for anisotropies in the point distribution.

In addition we have investigated the so-called freezing transition for
the partitions and for all finite subgroups of SU$(2)$. The main
result visualised in \cref{fig:fibPhaseScan} is that the partitioning
$F_k$ based on Fibonacci lattices allows for a flexible choice of the
number of elements by adjusting $k$ and at the same time larger
$\beta_c$-values compared to finite subgroups and the other discussed
partitionings. Thus, Fibonacci based discretisations provide the
largest simulatable $\beta$-range at fixed $n$.

Coming back to the introduction, using the partitionings proposed here
does not pose any problem even at very large $\beta$-values at least
in Monte Carlo simulations. This leaves us optimistic for their
applicability in the Hamiltonian formalism for tensor network or
quantum computing applications.

\begin{acknowledgments}
	\noindent This work is supported by the Deutsche
	Forschungsgemeinschaft (DFG, German Research Foundation) and the
	NSFC through the funds provided to the Sino-German
	Collaborative Research Center CRC 110 “Symmetries
	and the Emergence of Structure in QCD” (DFG Project-ID 196253076 -
	TRR 110, NSFC Grant No.~12070131001) as well as the STFC Consolidated Grant ST/T000988/1.
	The open source software packages R~\cite{R:2019} and
	hadron~\cite{hadron:2020} have been used.
\end{acknowledgments}

\bibliographystyle{h-physrev5}
\bibliography{bibliography}

\end{document}